\documentclass[symmetry,review,accept,moreauthors,pdftex]{mdpi}

\firstpage{1} 
\pubvolume{13}
\issuenum{12}
\articlenumber{2418}
\pubyear{2021}
\copyrightyear{2021}
\externaleditor{Academic Editors: Martiros Khurshudyan and Asatur Khurshudyan} 
\datereceived{15 November 2021} 
\dateaccepted{10 December 2021} 
\datepublished{14 December 2021} 
\hreflink{https://doi.org/10.3390/\linebreak sym13122418} 
\usepackage{mathtools}
\usepackage{url}
\def\UrlBreaks{\do\/\do-}
\Title{Principles of Gravitational-Wave Detection with Pulsar Timing~Arrays}
\TitleCitation{Principles of Gravitational-Wave Detection with Pulsar Timing Arrays}

\Author{Michele Maiorano $^{1,2,3,}$*\orcidA{}, Francesco De Paolis $^{1,2,3}$\orcidB{} and Achille A. Nucita $^{1,2,3}$\orcidC{}}
\AuthorNames{Michele Maiorano, Francesco De Paolis and Achille A. Nucita}
\AuthorCitation{Maiorano, M.; De Paolis, F.; Nucita, A.A.}
\address{
$^{1}$\quad Department of Mathematics and Physics ``Ennio De Giorgi'', University of Salento, Via Arnesano, \mbox{I-73100 Lecce, Italy;} depaolis@le.infn.it (F.D.P.); achille.nucita@le.infn.it (A.A.N.)\\
$^{2}$ \quad National Institute for Nuclear Physics, National Institute for Nuclear Physics (INFN), Sezione di Lecce, \mbox{Via Arnesano, I-73100 Lecce, Italy}\\
$^{3}$ \quad National Institute for Astrophysics, National Institute of Astrophysics (INAF), Sezione di Lecce, \mbox{Via Arnesano, I-73100 Lecce, Italy}}
\corres{Correspondence: michele.maiorano@le.infn.it}
\abstract{Pulsar timing uses the highly stable pulsar spin period to investigate many astrophysical topics. In particular, pulsar timing arrays make use of a set of extremely well-timed pulsars and their time correlations as a challenging detector of gravitational waves. It turns out that pulsar timing arrays are particularly sensitive to ultra-low-frequency gravitational waves, which makes them complementary to other gravitational-wave detectors. Here, we summarize the basics, focusing especially on supermassive black-hole binaries and cosmic strings, which have the potential to form a stochastic gravitational-wave background in the pulsar timing array detection band, and the scientific goals on this challenging topic. We also briefly outline the recent interesting results of the main pulsar timing array collaborations, which have found strong evidence of a common-spectrum process compatible with a stochastic gravitational-wave background and mention some new perspectives that are  particularly interesting in view of the forthcoming radio observatories such as the Five hundred-meter Aperture Spherical Telescope, the MeerKAT telescope, and the Square Kilometer Array.}

\begin{document}
\section{Introduction}
\label{introduction}

Gravitational waves (GWs), predicted by Albert Einstein's theory of General Relativity (GR)~\cite{einstein1916}, are ripples in space-time propagating at the speed of light in a vacuum, emitted by compact object systems characterized by a quadrupole moment with a non-null second time derivative. Since the strain associated with GWs is proportional to their amplitude $h$, which is in most cases less than $ 10^{-16}$, even Einstein himself wondered if they could ever be discovered~\cite{cervantes2016}. Over the years, several different techniques have been developed, up to the first successful event detection, named GW150914~\cite{abbott2016}, emitted during a black-hole (BH) merger, which was achieved in 2015 by the LIGO (Laser Interferometer Gravitational-Wave Observatory) collaboration, thanks to the Livingston and Hanford gravitational interferometers. Then,  a new window for observational astronomy opened up, indicating, especially with the discovery of the first binary neutron star merging event (GW170817)~\cite{abbott2017} observed alongside its electromagnetic counterpart, i.e., a Short Gamma-Ray Burst (GRB170817)~\cite{goldstein2017} detected by the Fermi satellite, the start of the multi-messenger era.

Ground-based laser interferometers, such as those of the LIGO and VIRGO collaborations, are sensitive only to high-frequency (i.e., in the frequency range $[10,10^4]$ Hz) GWs, so the main detectable GW sources are binary black holes (BBH) or binary neutron stars (BNS) in the final stage of their evolution. Third-generation ground-based laser interferometers, such as ET (Einstein Telescope) and CE (Cosmic Explorer), planned for the near future, will be even more sensitive (i.e., in the frequency range $[1,10^4]$ Hz). However, to observe compact object systems in earlier phases, alongside other types of low-frequency (i.e., in the frequency range $[10^{-5},1]$ Hz) GW sources, such as extreme mass-ratio inspirals (EMRI) and merging supermassive black-hole binaries (SMBHBs), space-based laser interferometers  that bypass the problems due to the Earth seismic noise  are needed. The most promising project in this direction is LISA (Laser Interferometer Space Antenna), which will consist of a constellation of three satellites, separated by about $2.5\times 10^6$ km, in a triangular configuration, flying around the Sun along and Earth-like heliocentric \mbox{orbit~\cite{lisa2017}}. Even if LISA, whose launch is planned in the next decade, is an extremely advanced detector, it will likely not be able to detect ultra-low-frequency (i.e., in the frequency range $[10^{-10},10^{-6}]$ Hz) GWs. These GWs are expected to be generated by many sources of cosmological interest, such as inspiralling SMBHBs~\cite{rajagopal1995} or cosmic strings~\cite{damour2001}. Detecting such GWs is possible, however, through   pulsar timing arrays (PTA), which exploit the telescopes generally used for radio astronomy to measure the very tiny variations in the times of arrival (ToA) of the pulses emitted by millisecond pulsars (MSP), induced by GWs. Eventually, it is worth remarking that PTAs and LISA can work synergically as they are both sensitive to the GW emissions from SMBHBs. Indeed, PTAs can detect the GW emissions from an SMBHB in the earlier inspiral phases, while LISA can detect the GW emissions from the same source but in the later merger and ring-down phases. Therefore, they may be used to investigate the evolution of SMBHBs. For an overview of the expected range of amplitude and frequency of GW emissions for different types of GW sources, the reader is referred to Refs.~\cite{moore2015,gwplotweb}.

This paper aims to review the science behind PTA, the basis of detection, the main scientific goals, and the current results.

\section{Pulsar Timing Arrays}
\label{pulsartimingarrays}
A PTA is a set of pulsars that are constantly monitored by several ground-based radio telescopes to collect the ToAs of pulses emitted by them. This procedure is called pulsar timing and can be used for many purposes, from searching for extra-solar planets~\cite{wolszczan1992} to hunting for ultra-low-frequency GWs~\cite{sazhin1978}. The main collaborations that are currently working on PTAs are the European Pulsar Timing Array (EPTA)~\cite{desvignes2016}, the Indian Pulsar Timing Array (InPTA)~\cite{joshi2018}, the North American Nanohertz Observatory for Gravitational Waves (NANOGrav)~\cite{arzoumanian2018}, and the Parkes Pulsar Timing Array (PPTA)~\cite{reardon2016}. They join their efforts as the International Pulsar Timing Array (IPTA)~\cite{verbiest2016}.

\subsection{Millisecond Pulsars}
At the moment, there are more than $3000$ known pulsars~\cite{atnfweb} (see also Ref.~\cite{manchester2005}). Unfortunately, the majority of these pulsars are not suitable for PTAs. The only pulsars that meet the criteria for the required stability  are MSPs. These objects are the most precise clocks that can be timed because the variation $\dot{P}$ of their spin period $P$, mainly caused by the emission of magnetic dipole radiation, is extremely small ($\dot{P}\lesssim 10^{-19}$ s s$^{-1}$)~\cite{duncan2008}. Of the approximately $400$ MSPs discovered until now~\cite{atnfweb}, only about $10\%$ of them are actually used in PTAs because the others are affected by a timing noise whose level is not good enough to allow reliable GW detection~\cite{arzoumanian2020a}. The PTA pulsars are mostly isolated and binary Galactic MSPs, except for J1824$-$2452A, which lies in the core of the $M28$ globular cluster (GC).

\subsection{Pulse Time of Arrivals}
The pulsar ToAs are not directly measurable but must be inferred by the observed pulse profile, which shows how the pulsar luminosity changes with the pulsar phase. However, single pulses are usually too weak to be detected one by one, and there is a large variability between each observed pulse profile. A common procedure used to increase the signal-to-noise ratio~\cite{lorimer2005} and to improve the detection stability is to produce the integrated pulse profile, taking the average of a series of pulses. This operation reduces the uncorrelated Gaussian noise, giving a more clean and stable pulse profile. The corresponding ToA is then obtained by cross-correlating it with a high signal-to-noise template, which is typically the sum of pulse profiles over many epochs~\cite{mclaughlin2016}.

The observed ToAs carry a lot of useful information, which can be extracted by comparing them to the predicted ToAs. To obtain a timing model, one must express the N-th rotation of the pulsar as Taylor series, with respect to time~\cite{backer1986}:
\begin{equation}
\label{nthrotation}
N=\nu t+\frac{1}{2}\dot{\nu}t^2+\frac{1}{6}\ddot{\nu}t^3+...
\end{equation}

Equation \eqref{nthrotation} can be inverted in order to obtain the expression of the N-th ToA:
\begin{equation}
\label{timeofarrival}
t=\nu^{-1}N-\frac{1}{2}\dot{\nu}\nu^{-3}N^2-\frac{1}{6}\ddot{\nu}\nu^{-4}N^3+...
\end{equation}

Note that Equation \eqref{timeofarrival} only holds at the position of the pulsar. However, the ToAs are measured with respect to the Earth, which is not a good reference frame because of its motion around its axis and around the Sun. Therefore, it is more convenient to use a reference frame centered in the Solar System barycenter (SSB) that, within this context, can be considered as inertial. The reference frame currently used for pulsar timing is the International Celestial Reference Frame (ICRF). This is a realization of the International Celestial Reference System (ICRS), which is the standard reference system adopted by the International Astronomical Union (IAU), centered in the SSB. Once the position of the SSB~\cite{vallisneri2020} is determined with sufficient precision, the timing model, in the general case of a binary pulsar, must include all the effects that can cause delays or advances in the \mbox{ToAs~\cite{foster1990}}:
\begin{equation}
\label{timeofarrivalcorrections}
t=t_{\oplus}-\Delta_{\odot}-\Delta_{ism}-\Delta_{bin}
\end{equation}

Here, the term $\Delta_{\odot}$ includes the effects pertinent to the transformation of the pulse ToAs at the Earth ($t_{\oplus}$) to the equivalent at the SSB, namely atmospheric delays, vacuum retardation due to observatory motion (Roemer delay and parallax), the dispersion delay due to the free electrons content of the solar wind, the effects of relativistic frame transformation (Einstein delay) and the delay experienced by photons as they go through the gravitational potential generated by  Solar System bodies (Shapiro delay). The $\Delta_{ism}$ term includes the effects due to the propagation time of the signal from the pulsar (or binary system barycenter) to the SSB, namely the vacuum propagation delay (the path length divided by the light speed in vacuum), the dispersion delay due to the free electron content of the interstellar medium and other frequency-dependent delays, and $\Delta_{bin}$ includes the effects caused by the presence of a binary companion, namely the excess in the vacuum light travel time due to the Euclidean displacement of the pulsar (binary Roemer delay), a pseudo-delay that accounts for the aberration of the radio beam by binary motion, the gravitational redshift and special relativistic time dilation in the pulsar frame (binary Einstein delay) and the gravitational time dilation in the proximity of the companion (binary Shapiro delay). Obviously, in the case of an isolated pulsar, all the terms associated with the binary companion can be set to zero. For an exhaustive compendium of all these effects and their mathematical expressions, the reader is referred to Ref.~\cite{edwards2006}.

The majority of the effects affecting ToAs are, in general, well modeled and, therefore, they can be removed with the available pulsar timing analysis software such as \mbox{Tempo2~\cite{hobbs2006,edwards2006}}, which is one among those most commonly used.  It follows that the most significant effects for pulsar timing analysis are those that cannot be  modeled due to their stochastic nature. This is the case, for example, of the dispersion delay $\Delta_{dis}$, which is given~by:
\begin{equation}
\label{eq:disdel}
\Delta_{dis} = \frac{e^2}{2\pi m_e}\frac{1}{\nu_{rad}^2}\int_0^{d_p}dx\ n_{e}
\end{equation}
where $e$ is the electron charge, $m_e$ is the electron mass, $\nu_{rad}$ is the MSP radio-emission frequency, $d_p$ is the MSP distance from the observer, and $n_e$ is the electron number density~\cite{lorimer2005,tiburzi2016}. For convenience, geometrical units $c=G=1$ have been adopted unless otherwise specified. As can be seen from Equation \eqref{eq:disdel}, the dispersion delay is subject to stochastic variations caused by the random interstellar medium density fluctuations.

\subsection{Timing Residuals}
The values of the timing model's parameters can be obtained by fitting the function that predicts the ToAs to the observed ones, including the corrections expressed by \mbox{Equation \eqref{timeofarrivalcorrections}}. The timing model has to be continuously updated to include newly discovered effects, as long as there are 
clock corrections, if any. If all known effects that can influence the ToAs have been taken into account, the timing residuals, calculated by the difference between the fitting function and the ToAs data-set, are expected to be uniformly distributed around zero. The presence of structures in the timing residuals can be, therefore, the signature of unmodeled phenomena. In particular, as will be discussed in Section \ref{gravitationalwavesdetection}, GWs can induce timing residuals in the pulsar ToAs, with a shape dependent on the GW source.

\section{Gravitational-Wave Detection}
\label{gravitationalwavesdetection}

The detection of GWs is essentially based on a GR result according to which their passage can ``stretch'' and ``compress'' the space-time. This implies that the time of flight of an electromagnetic signal propagating from one position to another is not a constant, but can be longer or shorter than it would have been in the absence of GWs. This principle, which has already been successfully applied to the detection of high-frequency GWs through gravitational interferometry, would also allow revealing ultra-low-frequency GWs by pulsar timing. The frequency range to which PTAs are sensitive is determined by the duration of the observation campaign and the cadence of observations. The current observation campaign is thirteen years long, and observations are performed mainly once per week. This leads to   sensitivity in the frequency range $[10^{-9},10^{-6}]$ Hz.

To explain how a GW can influence the ToA of a pulse, let us first consider an ideal situation where an observer is located in the SSB, which is the origin of a right-handed three-dimensional reference frame, and a MSP located at a distance $d_p$ from the observer. As next, let us consider a GW emitted by a source at distance $d_s$ (with $d_s\gg d_p$), which goes through this two-body system, ignoring for the moment the presence of other effects which will be discussed in Section \ref{pulsartimingarrays}. This allows us to use the plane-wave approximation, thereby writing the GW as a superposition of its polarization states. Since the GW amplitude is small, the space-time metric tensor $g_{\mu\nu}$ can be written as the sum of the Minkowski tensor $\eta_{\mu\nu}$ and the perturbation tensor $h_{\mu\nu}$, with $\lvert h_{\mu\nu}\rvert\ll\lvert\eta_{\mu\nu}\rvert$. Writing down the squared distance $ds^2$ between two space-time events and considering that for light-type events $ds^2=0$ one obtains a differential equation that, if integrated along the unperturbed light path for two consecutive MSP pulses, gives the period variation that can be measured by the observer. If the pulsar period is much less than the time of flight of the GW period, a condition that especially holds for ultra-low-frequency GWs, one has:\vspace{-3pt}
\begin{equation}
\frac{\Delta P}{P}=\frac{p^ip^j}{2}\int_{t_e}^{t_e+t_p}dt \ \frac{\partial h_{ij}(t,x^i)}{\partial t}\vspace{-3pt}
\label{deltatau}
\end{equation}
where $p^i$ is the versor in the direction of the pulsar, $t_e$ is the time of the emission of the first pulse, $t_p$ is the light travel time between the pulsar and the observer and the perturbation tensor depends on the time $t$ and the position $x^i$. From Equation \eqref{deltatau}, writing the tensor $h_{ij}$ as the superposition of the ``plus'' ($+$) and ``cross'' ($\times$) polarization states of the GW, and making explicit the dependencies of the amplitude $h^A$ of the $A$-th polarization state as $h^A=h^A(-k_\mu x^\mu)$, where $k^\mu=(\omega,\vec{k})$ is the wavenumber four-vector of the GW, one \mbox{finds~\cite{maggiore2008a}:}\vspace{-3pt}
\begin{equation}
z(t,\Omega^i)=\sum_{A=+,\times}F^A[h^A(t,x^i=0)-h^A(t-t_p,x^i=t_pp^i)]\vspace{-3pt}
\label{redshift}
\end{equation}
where $z=-\Delta P/P$ is the observed relative period variation, the superscript $A$ indicates the $A$-th polarization, and $F^A$ is the antenna pattern function, given by:\vspace{-3pt}
\begin{equation}
\label{antennapattern}
F^A=\frac{1}{2}\frac{p^ip^je_{ij}^A}{(1+\Omega_ip^i)}
\end{equation}
where $e_{ij}^A$ is the polarization tensor, and $\Omega^i$ is the versor in the direction of the GW emission. In Equation \eqref{redshift}, the two terms in the square brackets are often called, respectively, the Earth term and pulsar term because the former indicates the metric perturbation in the proximity of the Earth (at SSB), while the latter refers to the pulsar.

The quantity $z$ will be here referred to as pulse redshift to distinguish it from the photon redshift~\cite{boitier2020}, which is the observed frequency variation of an electromagnetic signal. Actually, it can be shown that a GW can also induce a photon redshift described by the same equation, but in this case, the detection technique is not based on pulsar \mbox{timing~\cite{soyuer2021}}. The GW-induced timing residuals can be defined by integrating  Equation \eqref{redshift} with respect to the time:\vspace{-3pt}
\begin{equation}
r(t,\Omega^i)=\int_0^tdt' \ z(t',\Omega^i)\vspace{-3pt}
\label{gravitationaltimingresidual}
\end{equation}

Equation \eqref{gravitationaltimingresidual} describes the observed variation of the ToAs of the pulses induced by the \mbox{GW~\cite{detweiler1979}}. The antenna pattern function, defined in Equation \eqref{antennapattern}, describes how the gravitational radiation is irradiated into space. This appears more clearly when the antenna pattern function is written explicitly. In the adopted reference frame, the $z$-axis is oriented in the direction of the GW source and the coordinates $\theta$ and $\phi$ indicate, respectively, the angle between the direction of the GW source and that of the MSP and the angle between the $x$-axis and the projection on the $xy$-plane of the direction of MSP. With these choices, one obtains  $\Omega^i=(0,0,1)$ and $p^i=(\sin\theta\cos\phi, \sin\theta\sin\phi, \cos\theta)$. Considering that in GR the polarization tensors are:\vspace{-3pt}
\begin{equation}
\label{polarizations}
e_{ij}^+=\begin{pmatrix}
1 &0 &0 \\ 
0 &-1 &0 \\ 
0 &0 &0 
\end{pmatrix}\quad
e_{ij}^\times=\begin{pmatrix}
0 &1 &0 \\ 
1 &0 &0 \\ 
0 &0 &0 \vspace{-3pt}
\end{pmatrix}
\end{equation}
and using the Weinberg sign convention for the Minkowski tensor, one has~\cite{yunes2013}:\vspace{-3pt}
\begin{equation}
\label{antennae}
F^+=\frac{1}{2}\frac{\sin^2\theta}{1+\cos\theta}\cos(2\phi)\quad F^\times=\frac{1}{2}\frac{\sin^2\theta}{1+\cos\theta}\sin(2\phi)\vspace{-3pt}
\end{equation}

\begin{figure}[H]
\includegraphics[scale=0.8]{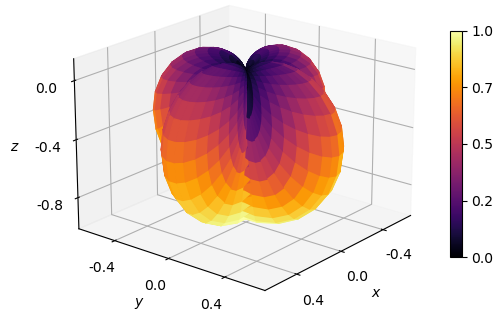}
\caption{Plot of the antenna pattern function for the $+$ polarization. The function is taken as the absolute value and is plotted in Cartesian coordinates $x$, $y$, and $z$. The $z$ axis is oriented along the direction of the GW source. The color indicates the magnitude of the function, as described by the color bar. The antenna pattern function for the $\times$ polarization has the same shape, but is rotated by $\Delta\phi={\pi}/{4}$.}
\label{antennaeplot}
\end{figure}

As can be noticed from Equation \eqref{antennae} and Figure \ref{antennaeplot}, the maximum of the antenna pattern function occurs for $\theta=\pi$, which is when the GW source and the MSP are aligned with the observer. Here, there is a removable discontinuity, that can be fixed by replacing $F^A(\pi)$ with $\lim_{\theta\to \pi}F^A$.

\section{Supermassive Black-Hole Binaries}

\subsection{Continuous Waves}
\label{continuousgravitationalwaves}
By now, it is well known that almost all bright galaxies harbor a supermassive black hole (SMBH)~\cite{kormendy2013}. Observations on several stars close to the center of the Milky Way, for example, confirmed the presence of a SMBH of about $4\times10^6$\,M$_\odot$~\cite{ghez2008,genzel2010}. So, massive black holes cannot be the remnant of stellar evolution; their formation likely started in the early Universe from smaller BH seeds that evolve through accretion and successive merging events. It is thought that BH seeds may be the remnant of Population III stars that, due to their low metallicity, could grow up to one hundred solar masses~\cite{madau2001}, or  else  are primordial BHs formed during the early stages of the Universe~\cite{carr1974}. It is also possible that BH seeds are formed directly by the collapse of large gas clouds~\cite{bonoli2014}. In any case, it is expected that the formation of SMBHBs occurs when two galaxies merge. In this scenario, the whole system loses energy due to dynamical friction, and this process reduces the separation between the two SMBHs at the center of the galaxy. Dynamical friction energy loss becomes less effective as the SMBHB orbital period shortens and eventually stops when it becomes smaller than the interaction time of the galaxy stars. At this stage, the GW emission is negligible, so in the absence of other energy subtraction processes, the SMBHB would stall indefinitely. This is known as the last parsec problem because it is still not clear which mechanism is responsible for the shrinking of the SMBHB down to sub-parsec scales, where the GW emission starts to dominate. Once this phase is reached, the SMBHB emits continuous GWs with a frequency of the order of $10^{-9}$ Hz. Such GWs induce a space-time perturbation that, within the context of GR, can be described by:
\begin{equation}
\label{continuousperturbation}
h^A=\operatorname{Re}\lbrace h_{cgw}^Ae^{i(k_\mu x^\mu+\alpha^A)}\rbrace
\end{equation}
where $i=\sqrt{-1}$ denotes the imaginary unit, $h_{cgw}^A$ and $\alpha^A$ are, respectively, the amplitude and the initial phase of the A-th polarization. Notice that it has been specified that one must take the real part (indicated by $\operatorname{Re}$) of the right member since $h^A$ is a physical quantity. Here, it is important to remark that the GW frequency is not a constant. It can be shown that for an SMBHB that has been circularized by GW emission, the emitted GW frequency is about two times the size of the orbital frequency $f_{orb}$ of the system. Hence, accounting for the cosmological redshift, the observed GW frequency $f$ is given by:
\begin{equation}
\label{frequencyredshift}
f\simeq\frac{2}{1+z}f_{orb}
\end{equation}

The energy loss due to GW emission makes the SMBHB shrink so that the frequency of the emitted GWs grows with time. It can be shown that the variation of the orbital frequency is given by~\cite{shapiro1983}:
\begin{equation}
\label{frequencyvariation}
\frac{df_{orb}}{dt}=\frac{96}{5}(2\pi)^{8/3}\mu M^{2/3}f_{orb}^{11/3}
\end{equation}
\textls[-15]{where $\mu$ is the reduced mass of the SMBHB and $M$ is the total mass of the binary. \mbox{Equation \eqref{frequencyvariation}}} implies that an SMBHB in the ultra-low-frequency GWs emission regime evolves very slowly. For this reason, the frequency of a GW that passes through the MSPs and the Earth can be considered, in good approximation, the same.

Combining Equations \eqref{redshift}, \eqref{gravitationaltimingresidual} and \eqref{continuousperturbation},  one obtains:
\begin{equation}
\begin{split}
r(t,\Omega^i)=&\sum_{A=+,\times}\operatorname{Re}\bigg\lbrace\frac{F^Ah_{cgw}^A}{-i\omega }[e^{-i\omega t+i\alpha^A}-e^{-i\omega[t-t_p(1+\cos\theta)]+i\alpha^A}-\\
&e^{i\alpha^A}+e^{i\omega t_p(1+\cos\theta)+i\alpha^A}]\bigg\rbrace
\end{split}
\label{residualcontinuous}
\end{equation}

\begin{figure}[H]
\includegraphics[scale=0.8]{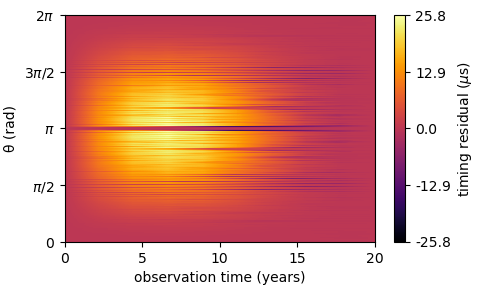}
\caption{Plot of the GW-induced timing residual function for a continuous GW. It has been considered a GW in the direction $\phi=0$, with an amplitude $h_{cgw}^+=10^{-16}$, a frequency $\omega=10^{-8}$ Hz, detected by a MSP at a distance of $d_p=1$ kpc. The horizontal axis indicates the observation time, expressed in years. The vertical axis indicates the angle $\theta$, expressed in radians. The color indicates the magnitude of the function, as described by the color bar.}
\label{colormap}
\end{figure}

\begin{figure}[H]
\includegraphics[scale=1]{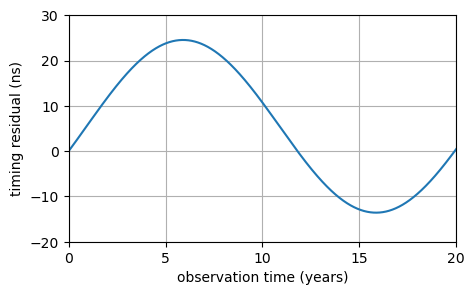}
\caption{Plot of the GW-induced timing residual function for a continuous GW. It has been considered a GW in the direction $(\theta,\phi)=(3,0)$, with  amplitude $h_{cgw}^+=10^{-16}$ and frequency $\omega=10^{-8}$ Hz, detected by a MSP at a distance of $d_p=1$ kpc. The horizontal axis indicates the observation time, expressed in years. The vertical axis indicates the timing residuals, expressed in nanoseconds.}
\label{resplot}
\end{figure}

The function in Equation \eqref{residualcontinuous} is periodic with respect to both variables $\theta$ and $t$ and, as can be seen from Figure \ref{colormap}, it rapidly oscillates with respect to $\theta$. Moreover, due to the GW transverse nature, it is null when the GW source and the MSP are aligned with the observer, and assumes its global maximum for $\theta\simeq\pi$. For a given value of $\theta$, as can be seen from Figure \ref{resplot}, the GW-induced timing residuals show a sinusoidal behavior. The GW angular frequency $\omega$ determines its period, while the factor $h_{cgw}^A/\omega$ gives an upper limit on timing residuals that can be induced by a GW emitted by an SMBHB. For $\omega\simeq 10^{-8}$ Hz, which is in the frequency range of GWs detectable by current PTAs, the typical value of the GW amplitude turns out to be about $10^{-16}$. Therefore, one has $h_{cgw}^A/\omega\simeq 10$ ns. Unfortunately, there are very few MSPs that can be timed with such high precision, so at the moment it is unlikely to detect single SMBHBs by pulsar timing.

\vspace{-12pt}

\subsection{Gravitational-Wave Background}
\label{stocasticgwb}
The ensemble of GWs emitted by a large number of independent sources is expected to form the so-called stochastic gravitational-wave background (GWB). Like the Cosmic Microwave Background (CMB), GWB carries a lot of information about the early Universe and possibly allows us to extend our observations beyond the last-scattering surface, from where the electromagnetic radiation could not escape. One of the largest contributions to GWB should be due to the population of SMBHBs. Since each of these BH binaries is in a different evolution stage and a different position, the pulse redshift associated with the ensemble can be obtained by integrating  Equation \eqref{redshift} with respect to frequency and direction. Thus, using the GW amplitude expressed in Equation \eqref{continuousperturbation}, this leads to:\vspace{-3pt}
\begin{equation}
z(t)=\sum_{A=+,\times}\int^\infty_{-\infty}df \ \int_{\mathcal{S}^2}d^2\Omega^i \ h_{cgw}^A(f,\Omega^i)F^A(\Omega)e^{-i\omega t}\left[1-e^{i\omega t_p(1+\Omega_ip^i)}\right]\vspace{-3pt}
\label{redshiftintegrated}
\end{equation}

\begin{figure}[H]
\includegraphics[scale=1]{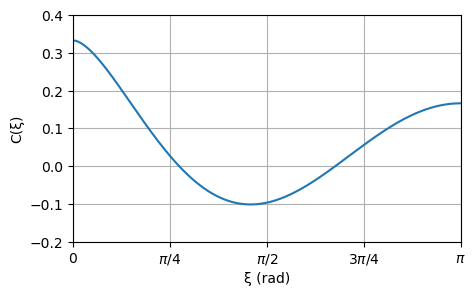}
\caption{Plot of the The Hellings and Downs function. The Hellings and Downs function $C(\xi)$ in Equation (\ref{hellingsanddowns}) is plotted as a function of the angular separation $\xi$ between the MSPs, expressed in radians. The function $C(\xi)$ indicates the cross correlation between the monitored MSPs expected in the case of the GWB.}
\label{hellings}
\end{figure}

A useful way to characterize statistically any stochastic background is by computing the ensemble average of its components, which are random variables. This quantity is equivalent to the time average only if the ergodic hypothesis holds. In the case of the GWB, some well-motivated assumptions can be made to simplify the calculation~\cite{maggiore2008b}. For a stationary, Gaussian, isotropic and unpolarized GWB, one has:\vspace{-3pt}
\begin{equation}
\left<h_{cgw}^{\ast A}(f,\Omega^i)h_{cgw}^{A'}(f',\Omega'^i)\right>=\delta(f-f')\frac{\delta^2(\Omega^i,\Omega'^i)}{4\pi}\delta^{A,A'}\frac{1}{2}S_h(f)\vspace{-3pt}
\label{gwbassumptions}
\end{equation}
where $S_h(f)$ is the spectral density of the GWB~\cite{maggiore2008a}. Equations \eqref{redshiftintegrated} and \eqref{gwbassumptions} allow us to determine the ensemble average of the pulse redshift of a pair of MSPs, labeled with $a$ and $b$ respectively, which is given by:\vspace{-3pt}
\begin{equation}
\left<z_a^\ast(t)z_b(t)\right>=\frac{1}{8\pi}\int^\infty_{-\infty}df \ S_h(f)\int_{\mathcal{S}^2}d^2\Omega^i \ \sum_{A=+,\times}F^A_a(\Omega^i)F^A_b(\Omega^i)\vspace{-3pt}
\label{ensembleaverage} 
\end{equation}
where the asterisk indicates the operation of complex conjugation. Notice that the term in the square bracket in Equation \eqref{redshiftintegrated}, does not appear in Equation \eqref{ensembleaverage}. Indeed, for MSPs typically used in PTAs, $t_p$ is larger with respect to the inverse of $\omega$ and, for this reason, the complex exponentials oscillate very rapidly with the frequency. Therefore, in this limit, their overall contribution to the integral can be neglected. The second integral in Equation \eqref{ensembleaverage} is the cross correlation between the antenna pattern function of the MSPs. The estimation of this integral can be performed by adopting a more convenient reference frame. If the $z$ axis is chosen to be aligned to one MSP, it results in:\vspace{-3pt}
\begin{equation}
\begin{split}
p^i_a =& (0,0,1) \\ 
p^i_b =& (\sin\xi,0,\cos\xi)\\
\Omega^i =& (\sin\theta\cos\phi,\sin\theta\sin\phi,\cos\theta) \\
m^i =& (\sin\phi,-\cos\phi,0)\\
n^i =& (\cos\theta\cos\phi,\cos\theta\sin\phi,-\sin\theta)\vspace{-3pt}
\end{split}
\label{versors}
\end{equation}
where $\xi$ is the angular separation between the MSPs  and $m^i$ and $n^i$ are orthogonal versors lying in the plane perpendicular to $\Omega_i$. Since the polarization tensors are given by:
\begin{equation}
\begin{split}
e_{ij}^+ =& m_im_j-n_in_j\\
e_{ij}^\times =& m_in_j+n_im_j
\end{split}
\label{generalpolarizations}
\end{equation}
the antenna pattern function can be written explicitly, in the adopted reference frame, using Equations \eqref{versors} and \eqref{generalpolarizations}. Then, from Equation \eqref{ensembleaverage} one obtains:\vspace{-3pt}
\begin{equation}
\left<z_a^\ast(t)z_b(t)\right>=C(\xi)\int^\infty_0df \ S_h(f)
\label{ensembleaveragecrosscorrelation} 
\end{equation}
where the term:
\begin{equation}
C(\xi)=\frac{1}{3}\left\lbrace 1+\frac{3}{2}(1-\cos\xi)\left[\ln\left(\frac{1-\cos\xi}{2}\right)-\frac{1}{6}\right]\right\rbrace
\label{hellingsanddowns}
\end{equation}
is known as the Hellings and Downs function~\cite{hellings1983}, which is plotted in Figure \ref{hellings}.

The ensemble average between the timing residuals of a pair of MSPs can be determined following the same approach. Since timing residuals are directly related to the ToAs of the pulses, which are   actually observable, this quantity is much more useful in searching for GWB evidence. From Equation \eqref{residualcontinuous}, one can derive:
\begin{equation}
\left<r_a^\ast(t)r_b(t)\right>=2C(\xi)\int^\infty_0df \ \frac{S_h(f)}{(2\pi f)^2}2\left[1-\cos(2\pi t)\right]
\label{residualsensemble}
\end{equation}

The Hellings and Downs function is considered the ``smoking gun'' of the GWB because it may allow the distinguishing of quadrupolar correlations, due to the GWs, from monopolar or dipolar correlations~\cite{tiburzi2016}, which can be due to the choice of reference clocks and to misplacement of the position of SSB, respectively.

It can be shown that the characteristic strain spectrum of the GWB is well approximated by a power law:
\begin{equation}
h_c(f)\propto f^\alpha
\label{characteristicstrain}
\end{equation}
where $h_c(f)$ is the characteristic strain amplitude, which is linked to the spectral density by $h^2_c(f)=2fS_h(f)$~\cite{maggiore2000}, and $\alpha$ is the spectral index which, in this case, is $\alpha=2/3$.

\subsection{Gravitational-Wave Bursts with and without Memory}
\label{burstwithmemory}
A sudden and temporary variation of the quadrupole moment of a system may lead to impulsive space-time perturbations, known as GW bursts. Possible sources of GW bursts are then SMBHs in hyperbolic or highly eccentric elliptic orbits~\cite{capozziello2008}. While in the former case there are multiple GW emissions, which mainly occur close to each periastron passage, in the latter case there is a single GW emission concentrated mostly at the distance of the closest approach; therefore, these events are much harder to detect. Even if PTAs are sensitive to GW bursts, they must last at least some weeks to be detectable, because of the limit imposed by the cadence of observations.

As a result of extremely energetic events, GW bursts may lead to a permanent space-time modification, and for this reason, these are referred to as GW bursts with memory (BWM). Possible sources of BWM are then core-collapse supernovae, where the BWM is produced by the asymmetric ejection of a large amount of matter and radiation~\cite{fryer2011}, or SMBHBs, where the BWM is associated with the strong and anisotropic GW emission in the final stage of the merger~\cite{seto2009}. The upper limit on the BWM strain, which can be estimated by the ratio between its Schwarzschild radius and its distance from the observer~\cite{braginskii1987}, turns out to be of the order of $10^{-22}$ for core-collapse supernovae and $10^{-15}$ for SMBHBs.

\begin{figure}[H]
\includegraphics[scale=1]{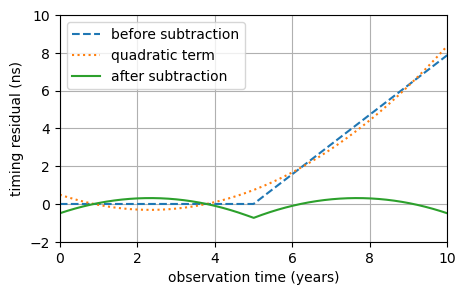}
\caption{Plot of the GW-induced timing residual function for a BWM. The dotted line describes the timing residuals due to the quadratic term, the dashed and the solid lines describe, respectively, the GW-induced timing residuals for a BWM before and after the quadratic term subtraction. It has been considered a BWM observed at a time $t_b=5$ yrs, during an observation campaign with a duration $t_{obs}=10$ yrs, in the direction $(\theta,\phi)=(\pi/2,0)$, with an amplitude $h_{bwm}=10^{-16}$. The axes indicate the observation time, expressed in years, and the timing residuals, expressed in nanoseconds.}
\label{bwm}
\end{figure}

Analogously to the case of continuous GWs,  BWM can also induce timing residuals. The space-time perturbation can be modeled as a step function:\vspace{-3pt}
\begin{equation}
h^+(t)=h_{bwm}\Theta(t-t_b)\quad h^\times(t)=0\vspace{-3pt}
\label{stepfunction}
\end{equation}
where $h_{bwm}$ is the BWM amplitude, $t_b$ is the observation time of the BWM and $\Theta(t-t_b)$ is the Heaviside function, which is null before the burst (i.e., for $t\leq t_b$) and it is unitary after it (i.e., for $t>t_b$). Taking into account Equations \eqref{redshift} and \eqref{stepfunction}, by integrating Equation \eqref{gravitationaltimingresidual} results~\cite{pshirkov2010}:\vspace{-3pt}
\begin{equation}
r(t,\Omega^i)=F^+h_{bwm}\left[(t-t_0)\Theta(t-t_0)-(t-t_1)\Theta(t-t_1)\right]\vspace{-3pt}
\label{residualbwm}
\end{equation}
where $t_0=t_b$ and $t_1=t_b-\omega t_p(1+\cos\theta)$, and therefore, the two terms in the right member are, respectively, the Earth term and the pulsar term. The BWM-induced timing residuals, expressed in Equation \eqref{residualbwm}, can be rewritten in a simpler form by neglecting the pulsar term because it gives rise to an uncorrelated contribution in the timing residuals of each MSP, and is then treated as a source of stochastic noise in the whole PTA \mbox{analysis~\cite{pshirkov2010,cordes2012}}. Moreover, it has to be considered that uncertainties in the MSP spin-period and spin-period first-derivative lead, respectively, to linear and quadratic contributions in the timing residuals~\cite{arzoumanian2015}. Therefore, after removing them, the GW-induced timing residuals for a BWM become:\vspace{-3pt}
\begin{equation}
r(t,\Omega^i)=F^+h_{bwm}\mathcal{I}(t)
\label{redidualbwmsubtracted}
\end{equation}
where $\mathcal{I}(t)=(t-t_b)\Theta(t-t_b)-\mathcal{I}_{qdr}(t)$ is the Earth term subtracted by the quadratic term $\mathcal{I}_{qdr}(t)=a(t-t_b)^2+b(t-t_b)+c$, plotted in Figure \ref{bwm}. The coefficients $a$, $b$ and $c$ are chosen to minimize the root mean square (RMS) of $\mathcal{I}(t)$.

\section{Cosmic Strings}
\label{cosmicstringsemission}

\subsection{Cosmological Origin of the Cosmic Strings}
\label{cosmicstrings}
According to the Cosmological Standard Model, the Universe began with an immediate expansion, referred to as the Big Bang, from an initial singularity characterized by extreme values of density and temperature. Right after the Big Bang,    an accelerated expansion epoch started called inflation, which lasted for about $10^{-32}$ s, during which the Universe's size increased due to the inflation field, and nowadays it is still increasing due to the dark energy, which became relevant about $10$ Gyr after the Big Bang. Since the Universe can be considered as an isolated system, its expansion is adiabatic and led to a decrease in temperature. Therefore, the Universe, at some point in its evolution, may have undergone a phase transition, during which some of the existing symmetries may have been broken and, as a consequence, topological defects might have arisen. In the case of the $U(1)$ symmetry, these structures are one dimensional and, for this reason, take the name of cosmic strings.

\subsection{Cosmic Strings from U(1) Spontaneous Symmetry Breaking}
\label{cosmicstringsuone}

A possible origin of the cosmic strings is the spontaneous symmetry breaking of a theory characterized by a U(1) global symmetry and Lagrangian density given by:
\begin{equation}
\mathcal{L}=\partial^*_\mu\phi\partial^\mu\phi-V(\phi)
\label{lagrangiandensity}
\end{equation}
where $\phi$ is the only scalar field of the theory and $V(\phi)$ is a Mexican hat potential:
\begin{equation}
V(\phi)=\frac{\lambda}{2}\left(|\phi|^2-\frac{1}{2}\eta^2\right)^2
\label{mexicanhatpotential}
\end{equation}
where $\lambda$ is a coupling constant and $\eta$ is the vacuum state of the field $\phi$, corresponding to the minimum of its potential $V(\phi)$. For convenience, in this section, natural units $c=\hbar=1$ have been adopted. The equations of motion for the density Lagrangian in Equation \eqref{lagrangiandensity}, which can be derived from the Euler--Lagrange equations, results:
\begin{equation}
\Box\phi=m_s^2\left(\frac{|\phi|^2}{\eta^2}-\frac{1}{2}\right)\phi
\label{equationofmotion}
\end{equation}
where $\Box=\partial_\mu\partial^\mu$ denotes the d'Alembert operator and $m_s=\sqrt{\lambda}\eta$. Equation \eqref{equationofmotion} can be solved by expressing the scalar field $\phi$ as:
\begin{equation}
\phi(\rho,\theta)=\frac{\eta}{\sqrt{2}}f(m_s\rho)e^{in\theta}
\label{complexscalarfieldpolarcoordinates}
\end{equation}
where the cylindrical coordinate system $(\rho,\theta,z)$ has been adopted, $f$ is a function and $n$ is an integer. From Equation \eqref{equationofmotion}, both a massless scalar field, associated with the spontaneous symmetry breaking (i.e., is the Goldstone boson) and a massive scalar field with mass $m_s$, arise. Since the function depends on the product between the mass of the scalar field $\phi$ and the radial distance, it is convenient to define $\xi=m_s\rho$. By substituting Equation \eqref{complexscalarfieldpolarcoordinates} in Equation \eqref{equationofmotion}, one has:
\begin{equation}
\frac{\partial^2f}{\partial\xi^2}+\frac{1}{\xi}\frac{\partial f}{\partial\xi}-\frac{n^2}{\xi^2}f-\frac{f}{2}(f^2-1)=0
\label{differentialequationstring}
\end{equation}

Equation \eqref{differentialequationstring} implies that the function $f$, and hence the scalar field $\phi$, has two different limits depending on $\xi$. For $\xi\rightarrow 0$, one must have $f\rightarrow 0$ to guarantee the validity of Equation \eqref{differentialequationstring}. For $\xi\gg 1$, instead, from Equation \eqref{differentialequationstring} results $f\rightarrow 1$. The energy density associated with the scalar field $\phi$ in Equation \eqref{complexscalarfieldpolarcoordinates} is:\vspace{-3pt}
\begin{equation}
\mathcal{E}=|\partial_\rho\phi|^2+\frac{1}{\rho^2}|\partial_\theta\phi|^2+V(\phi)
\label{energydensity}
\end{equation}

The function $f$ determines the behavior of the energy density in Equation \eqref{energydensity}. Indeed, for $\rho\gg m_s^{-1}$, one has $\mathcal{E}\propto \rho^{-2}$. The energy density is hence confined within a distance $\rho\simeq m_s^{-1}$, and since this is true $\forall z\in\mathbb{R}$, that defines a string~\cite{maggiore2008b}. The integer $n$ can be, therefore, interpreted as the winding number of the string. The energy density can be integrated over the element of surface $d\sigma =\rho d\rho d\theta$, to obtain the string tension $\mu$. Since for $\rho\rightarrow\infty$ the integral diverges, it must be performed within a cutoff radius $r_{cut}$:
\begin{equation}
\mu=\pi n^2\eta^2\log(m_sr_{cut})
\label{stringtension}
\end{equation}

A different origin of the cosmic strings is the spontaneous symmetry breaking of a theory characterized by a U(1) local symmetry and an Abelian--Higgs Lagrangian density, given by:
\begin{equation}
\mathcal{L}=-\frac{1}{4}F_{\mu\nu}F^{\mu\nu}+|D_\mu\phi|^2-V(\phi)
\label{abelianhiggslagrangian}
\end{equation}
where $F_{\mu\nu}=\partial_\mu A_\nu-\partial_\nu A_\mu$ is the curl of the vector field $A_\mu$, $D_\mu=\partial_\mu+igA_\mu$ is the covariant derivative operator, $g$ is a coupling constant, and $V(\phi)$ is the same Mexican hat potential introduced above (see Equation \eqref{mexicanhatpotential}). Therefore, in addition to a massive scalar field with mass $m_s=\sqrt{\lambda}\eta$, which is present also in the U(1) global symmetry, there is a massive vector field with mass $m_v=g\eta$, while is absent the massless scalar field, which is absorbed by the vector field and it is responsible for its mass. As for the U(1) global symmetry, the equation of motion can be derived from the Lagrangian density in Equation \eqref{abelianhiggslagrangian}, and it can be eventually solved by expressing the scalar and the vector fields as:
\begin{equation}
\begin{split}
\phi =& \eta f(m_s\rho)e^{in\theta}\\
A^i =& \frac{n}{g\rho}\hat{\theta}^ia(m_v\rho)
\end{split}
\label{fields}
\end{equation}
where $f$ and $a$ are two functions and $\hat{\theta}^i$ is the versor in the direction of the vector field. The solutions of the equation of motion are not trivial, but it can be found that asymptotically $f\rightarrow 1$ and $a\rightarrow 1$. Again, it can be shown that the energy density is localized in a \mbox{string~\cite{hindmarsh1995}}. The main difference with the U(1) global symmetry is that, for $\rho\rightarrow\infty$, the integral is finite, and the string tension, for $n=1$, is:
\begin{equation}
\mu=\pi\eta^2g(\beta)
\label{strintensionlocal}
\end{equation}
where $\beta=m_v/m_s$, and $g(\beta)$ is a slowly varying function that satisfies the condition $g(1)\rightarrow 1$.

\subsection{Gravitational-Wave Emission from Cosmic Strings}
\label{gravitationalwavesfromcosmicstrings}
Frequently, cosmic strings interact with each other or with themselves. From these processes, new structures, such as loops and kinks, might arise. Loops are cosmic strings in a ring-like configuration rapidly oscillating due to the high string tension and radiate energy via gravitational waves thereby shrinking and, eventually, decaying. During loop oscillations, the string can double back on itself, producing a cusp~\cite{blanco1999}. Kinks are cosmic strings in an open configuration characterized by a vertex. GW emission from cosmic strings might form a GWB. For a string network, the characteristic strain spectrum of the GWB can be described by a power law with a spectral index $\alpha=-7/6$~\cite{olmez2010}.

\begin{figure}[H]
\includegraphics[scale=1]{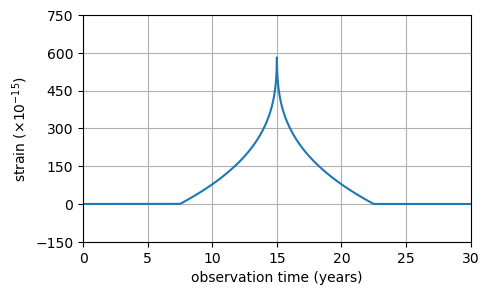}
\caption{Plot of the space-time perturbation associated with a GW burst emitted by a cusp. It has been considered a cosmic string at $z=1000$, with length $l=10$~kpc and tension $G\mu\simeq 10^{-6}$, emitting a GW burst with duration $w=15$ years and reaching peak at $t_0=15$ years. The horizontal axis indicates the observation time, expressed in years. The vertical axis indicates the strain, expressed in units of $10^{-15}$.}
\label{cuspamp}
\end{figure}

\begin{figure}[H]
\includegraphics[scale=1]{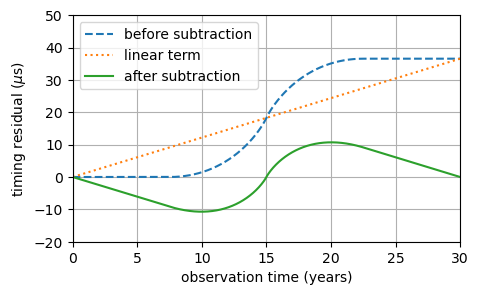}
\caption{Plot of the GW-induced timing residual function for a GW burst due to a cusp. The dotted line describes the timing residuals due to the linear term, the dashed and the solid lines describe, respectively, the GW-induced timing residuals for a GW burst due to a cusp before and after the linear term subtraction, represented by the dotted line.   A cosmic string at $z=1000$, with length $l=10$~kpc and tension $G\mu\simeq 10^{-6}$, emitting a GW burst in the direction $(\theta,\phi)=(\pi/2,0)$ with duration $w=15$ years has been considered, reaching peak at $t_0=15$ years. The horizontal axis indicates the observation time, expressed in years. The vertical axis indicates the timing residuals, expressed in $\mu$s.}
\label{cuspres}
\end{figure}

Cusps and kinks might emit GW bursts strong enough to stand above the GWB. In particular, cusps are expected to be powerful GW burst sources~\cite{damour2001}. The correspondent space-time perturbation can be written as:
\begin{equation}
h(t)^+=h_{csp}\Theta\left(t-t_0+\frac{w}{2}\right)\Theta\left(-t+t_0+\frac{w}{2}\right)\left[|t-t_0|^{1/3}\left(\frac{w}{2}\right)^{-1/3}-1\right]\,,\,\,\, h(t)^\times=0
\label{cuspsstrain}
\end{equation}
where $w$ is the duration of the GW burst emitted by the cusp, $t_0$ is the time relative to its peak, and $h_{csb}$ is its amplitude, given by:
\begin{equation}
h_{csb}=\sqrt{\frac{3}{2\pi}}\Gamma\left(-\frac{1}{3}\right)\left[\frac{1}{(1+z)}\frac{w}{2}\right]^{1/3}\frac{G\mu l^{2/3}}{r(z)}
\label{cuspsstrainamplitude}
\end{equation}
\textls[-15]{where $\Gamma$ is the Euler function, $l$ is the string length, and $r(z)$ is the comoving distance at redshift $z$~\cite{damour2001} (see Figure \ref{cuspamp}). The string length places a limit on GW frequency $f$, indeed, since the wavelength of the emitted GW cannot be greater than the string length, $f\gtrsim l^{-1}$. Therefore, PTAs can be used to detect GW burst from cosmic string with a length in the range $[10^{14}, 10^{17}]$ m. Within the context of Grand Unified Theory, cosmic strings should be characterized by a string tension $\mu\simeq 10^{-6}G^{-1}$ (i.e., $\mu\lesssim 10^{22}$ kg m$^{-1}$)~\cite{sakellariadou2007}. This is a huge value, just think that a cosmic string with a length of $10$ pc should have a mass of $10^9$ M$_\odot$ concentrated in a narrow filament, with a thickness, determined by $m_s^{-1}$ and $m_v^{-1}$ (see \mbox{Section \ref{cosmicstringsuone}}), that can be smaller than the proton radius~\cite{maggiore2008b}.}
The GW-induced timing residuals for a GW burst due to a cusp can be found using Equations \eqref{antennapattern}, \eqref{gravitationaltimingresidual}, \eqref{cuspsstrain} and \eqref{cuspsstrainamplitude}~\cite{yonemaru2021}. For simplicity, the pulsar term has been neglected, as was the case in  in Section \ref{burstwithmemory}. Due to the presence of the Heaviside function in	 \mbox{Equation \eqref{cuspsstrain}}, the result is a piecewise function expressed by:
\begin{equation}
 r(t,\Omega^i) = F^+(t,\Omega^i)\times
\begin{cases}
0\quad&\text{if $t<t_0-\frac{w}{2}$} \\
h_{csb}\bigg\lbrace\frac{3}{4}\left[\left(\frac{w}{2}\right)^{4/3}-|t-t_0|^{4/3}\right]+&\\
\text{sign}(t-t_0)\left(\frac{w}{2}\right)^{1/3}\left[t-t_0+\frac{w}{2}\right]\bigg\rbrace\quad &\text{if $t_0-\frac{w}{2}\leq t<t_0+\frac{w}{2}$}\\
h_{csb}\left(-\frac{1}{4}\right)\left(\frac{1}{2}\right)^{1/3}w^{4/3}\quad &\text{if $t\geq t_0+\frac{w}{2}$}
\end{cases}
\label{cusptimingresidual}
\end{equation}
where sign$(t-t_0)$ is the sign function, which is positive if the argument is greater than zero and negative otherwise. Moreover, as explained in Section \ref{burstwithmemory}, the linear contribution in the timing residuals can be subtracted to remove the uncertainties in the MSP period (see Figure \ref{cuspres}).

\section{Current Results}
\label{currentresults}
Recently, significant steps towards the detection of ultra-low-frequency GWs by PTAs have been made. All the main PTA collaborations might have observed a common-spectrum process compatible with a GWB, described by a power law characterized by a spectral index $\alpha=-2/3$ and a characteristic strain amplitude $h_c\simeq 10^{-15}$ at a frequency $f=1$ year$^{-1}$, produced either by a population of SMBHBs or by a cosmic string \mbox{network~\cite{blasi2021,lentati2015,goncharov2021}}. However, it is important to remark that, according to the NANOGrav analysis, the no-correlations hypothesis is rejected only mildly, with a $2\sigma$ significance, which is lower than the $3\sigma$ and $5\sigma$ significance standards of particle physics, required to claim for evidence and detection, respectively. Note that this result does not refer to time correlation, but instead to spatial correlation. The latter, as illustrated in Section \ref{stocasticgwb}, must show the Hellings and Downs signature (i.e., it must be a quadrupolar correlation) in order to claim a GWB detection consistent with GR. For detailed information about the observed common-spectrum process and complete data analysis of the most recent data sets released by EPTA, NANOGrav, and PPTA collaborations, the reader is referred to Refs.~\cite{chen2021,arzoumanian2020a,goncharov2021}.

In addition to SMBHBs and cosmic strings, there might be other sources of a common-spectrum process, which are worth considering. Some of the most suggestive and physically impactful, if confirmed, are inflation and reheating~\cite{guzzetti2016}, BBHs from the previous aeon merged during the Big Crunch~\cite{gorkavyi2022,rovelli2018} in cyclic conformal cosmological models of the Universe~\cite{penrose2010}, massive gravity~\cite{liang2021}, and axion-like particles~\cite{ratzinger2021}. In particular, a GWB produced by quantum fluctuations of the gravitational field during the inflationary epoch is actually predicted within the context of the standard cosmological model $\Lambda$-Cold Dark Matter ($\Lambda$CDM), so its discovery would be crucial for probing the evolution of the Universe. For a complete and detailed description of the inflationary epoch characteristic imprint expected on the observations and the upper limits that can be placed by PTAs, the reader is referred to Ref.~\cite{guzzetti2016}.

The GW nature of the common-spectrum process can be proved only by observing the Hellings and Downs signature in the cross-correlated timing residuals. While the GW detection cannot be claimed with the data currently available, significant constraints have been placed. For example, assuming a GWB produced by a population of SMBHBs, the merger rate has been constrained in the range $[1.2,45]\times 10^{-5}$ Mpc$^{-3}$ Gyr$^{-1}$ and has been placed a lower limit on the merger timescale, which is $\gtrsim 2.7$ Gyr. According to these results, the formation and the merging of a significant number of SMBHBs would be allowed~\cite{middleton2021}. Assuming instead a GWB produced by a cosmic string network, the string tension has been constrained in the range $[4,10]\times 10^{-11}\,G^{-1}$~\cite{ellis2021}. This improved by almost four orders of magnitude the upper limit placed by the Planck data on the \mbox{CMB~\cite{planck2016}}. Cosmic strings with such a low string tension would not emit GW bursts detectable by ground-based interferometers or PTAs~\cite{blanco2021}. It is worth mentioning that some very recent analyzes have highlighted the existence of some ``tension'' between the estimates on the string tension resulting from the data collected respectively by the NANOgrav and PPTA collaboration. In fact, while both could be compatible with a stochastic gravitational-wave background produced by cosmic strings, the string tension is constrained to be in the range $[2,30]\times 10^{-11}\,G^{-1}$ by the NANOgrav data and $\lesssim 4\times 10^{-11}G\,^{-1}$ by the PPTA Collaboration. However, it has been recently shown that this tension can be somewhat relaxed by invoking primordial Coleman--Weinberg inflation which partially inflates the string network~\cite{lazarides2021}.

PTA data can also be used in multi-messenger astronomy to confirm the presence of an SMBHB (in galaxies centers), showing a periodicity in the electromagnetic observations. Although an unambiguous GW detection has not yet been accomplished, PTAs can be used to constrain the physical parameters of a possible SMBHB hosted by a galaxy. For instance, one of the most intriguing galaxies suspected of hosting an SMBHB within its core is \mbox{$3$C $66$B~\cite{depaolis2004}}. The early constraint on its chirp mass, obtained using long-baseline interferometry, has been recently improved by almost one order of magnitude using PTA data, placing an upper limit $\mathcal{M}\lesssim (1.65 \pm 0.02)\times 10^9$ M$_\odot$~\cite{arzoumanian2020b}. This is still compatible with the SMBHB hypothesis, and hopefully, soon, PTAs will be able to either confirm or rule it~out.

\section{Conclusions and Further Perspectives}
\label{conclusions}

PTAs are very promising tools for ultra-low-frequency GW detection. They can be used to observe sources, such as SMBHBs and cosmic strings, by detecting continuous GW, GW bursts, and the GWB. This would allow addressing some extremely relevant astrophysical and cosmological questions about the Universe. Although PTAs already give interesting constraints on the GW sources above (see Section \ref{currentresults}), no GW detection has been claimed yet. However, the latest results of the PTA major collaborations suggest that the first GW detection is not that far from present times. Next-generation radio telescopes, as the Five hundred-meter Aperture Spherical Telescope (FAST)~\cite{nan2011}, the MeerKAT radio telescope~\cite{bailes2020}, and the Square Kilometre Array (SKA)~\cite{weltman2020}, will improve the timing precision. Therefore, in the next decade, the PTAs will probably be used as ultra-low-frequency GW~detectors. 

Eventually, it has to be said that even with better detectors and larger data sets, the possibility of having no GW detection cannot be excluded. There may be several explanations behind that case. One possibility is that there is not any signal to detect. For example, the last parsec problem (see Section \ref{continuousgravitationalwaves}) might prevent the merging of the SMBHBs, and cosmic strings (see Section \ref{cosmicstringsemission}) might not exist. Another possibility is that the GW detection method currently adopted by PTA, mainly based on the Hellings and Downs cross correlation, needs to be revised or supplemented. 

One way to do this could be, for example, by including MSPs inside the core of GCs in PTAs. Indeed, since these MSPs are confined in such a tiny angular region, which radius can be of the order of some arcseconds, the angular separation between them is so small that they should have strongly correlated GW-induced timing residuals (see Figure \ref{hellings}). This property, once all the possible ToAs variations due to the GC have been taken into account in the timing model, makes them very useful for GW detection. For a detailed discussion on this novel and important issue, the reader is referred to Ref.~\cite{maiorano2021}.

\vspace{6pt} 
\authorcontributions{M.M., F.D.P. and A.A.N. contributed equally to this work. All authors have read and agreed to the published version of the manuscript.}

\funding{This research received no external funding.}

\institutionalreview{Not applicable.}
\informedconsent{Not applicable.}
\dataavailability{No observational data have been used.} 
\acknowledgments{We acknowledge the support of the Theoretical Astroparticle Physics (TAsP) and Euclid projects of INFN. Andrea Possenti is acknowledged for useful discussions.}
\conflictsofinterest{The authors declare no conflict of interest.}
\abbreviations{Abbreviations}{The following abbreviations, reported in order of appearance, are used in this manuscript:\\

\noindent 
\begin{tabular}{@{}ll}
MDPI & Multidisciplinary Digital Publishing Institute\\
INFN & Istituto Nazionale di Fisica Nucleare\\
INAF & Istituto Nazionale di Astrofisica\\
GW & Gravitational Wave\\
GR & General Relativity\\
BH & Black Hole\\
LIGO & Laser Interferometer Gravitational-Wave Observatory\\
BBH & Binary Black Hole\\
BNS & Binary Neutron Star\\
ET & Einstein Telescope\\
CE & Cosmic Explorer\\
EMRI & Extreme Mass-Ratio Inspiral\\
LISA & Laser Interferometer Space Antenna\\
SMBHB & Supermassive Black-Hole Binary\\
PTA & Pulsar Timing Array\\
ToA & Time of Arrival\\
MSP & Millisecond Pulsar\\
EPTA & European Pulsar Timing Array\\
InPTA & Indian Pulsar Timing Array\\
NANOGrav & North American Nanohertz Observatory for Gravitational Waves\\
PPTA & Parkes Pulsar Timing Array\\
IPTA & International Pulsar Timing Array\\
GC & Globular Cluster\\
SSB & Solar System Barycenter\\
ICRF & International Celestial Reference Frame\\
ICRS & International Celestial Reference System\\
IAU & International Astronomical Union\\
SMBH & Supermassive Black Hole\\
GWB & Gravitational-Wave Background\\
CMB & Cosmic Microwave Background\\
BWM & Burst With Memory\\
RMS & root mean square\\
$\Lambda$CDM & $\Lambda$-Cold Dark Matter\\
FAST & Five hundred meter Aperture Spherical Telescope\\
SKA & Square Kilometre Array\\
TAsP & Theoretical Astroparticle Physics\\
\end{tabular}}
\newpage
\end{paracol}
\reftitle{References}

\end{document}